\documentclass{moriond}
\usepackage{amssymb}

\def\mco{\multicolumn}

\def\be{\begin{equation}}
\def\ee{\end{equation}}
\def\bea{\begin{eqnarray}}
\def\eea{\end{eqnarray}}

\begin{document}
\vspace*{4cm}
\title{ON THE DETECTABILITY OF BL LAC OBJECTS BY ICECUBE}

\author{ C. RIGHI$^{1,2}$, F. TAVECCHIO$^2$}

\address{$^1$Universit\'a degli studi dell'Insubria, DiSAT, Via Valleggio, 11 - 22100 Como, Italy\\
$^2$INAF - Osservatorio Astronomico di Brera, Via E. Bianchi 46, I-23807 Merate, Italy}

\maketitle\abstracts{
Since 2010 IceCube observed around 50 high-energy neutrino events of cosmic origin above 60 TeV, but the astrophysical sources of these events are still unknown. We recently proposed high-energy emitting BL Lac (HBL) objects as candidate emitters of high-energy neutrinos. Assuming a direct proportionality between high-energy gamma-ray and very-high energy neutrino fluxes, we calculated the expected neutrino event number in a year for IceCube and the presently under construction Km3NeT. To give a value of the significance of a detection we considered also the background for the single sources.  To this aim we derived the through-going muon rate, generated by muon neutrino including the effect of Earth absorption, the density of the Earth and the cross section $\nu N$. Applying this calculation both to HBL sources and the atmospherical neutrino background, we can calculate the expected significance of the detection by IceCube, showing that our scenario is compatible with a no detection of HBL.}

\section{Introduction}
In 2010 Ice Cube started to reveal neutrinos in clear excess to the expected atmospheric flux at very-high energy ($\gtrsim100$ TeV); this marked the beginning of the neutrino astrophysics era \cite{Aartsen13} \cite{Aartsen14}. Ice Cube is able to reconstruct some of the most relevant quantities (energy, $E_\nu$, and direction, $\alpha_\nu$ and $\delta_\nu$) of the incoming neutrino. There are two types of events: the high-energy starting events (HESE, or contained-vertex event) with a high angular uncertainty ($>1^\circ$) and the high-energy through-going muons, produced only by $\nu_\mu$ (and $\bar{\nu}_\mu$) \footnote{We will not distinguish from neutrino and antineutrino.} and with a good angular uncertainty ($\le 1^\circ$). Today we have $\approx 50$ events of different neutrino flavours\footnote{The tau neutrino $\nu_\tau$ is not been yet observed.} at these energies ($60$ TeV - $2.8$ PeV), but the cosmic source class (or classes) of these neutrinos is (are) still unknown.

 High-energy neutrinos are expected to be produced in regions rich of cosmic rays, where pions production is possible.  The production of PeV neutrinos requires very high energy protons interacting with matter or with photons\footnote{For $E_p \sim 10^{17}$ eV the required photon energy $E_\gamma$ is in the UV-X-ray range. }  to produce pions ($p+p \rightarrow X + \pi$ or $p+\gamma \rightarrow X + \pi$). Charged pions, in turn, decay in muon and neutrinos ($\pi^\pm \rightarrow \mu^\pm + \nu_\mu \rightarrow e^\pm + 2\nu_\mu + \nu_e$). The $p+p$ reaction could take place in regions with high barion density (such as galactic regions or star forming regions); meanwhile the photo-meson reaction is favoured in case of high photon density regions. After 5 years of data taking the sky distribution of the detected events is consistent with isotropy. The lack of a strong anisotropy suggests that the sources are not only galactic, but a mixed galactic and extragalactic origin can not be excluded. Recent studies suggest a possible contribution at ``low energy" ($30-100$ TeV) by galactic sources and an extragalactic component emission above $100$ TeV \cite{pall1}. Among the possible extragalactic sources there are star forming galaxies \cite{tamb1} \cite{Chakraborty}, active galactic nuclei (AGN) \cite{Kalashev} \cite{Jacobsen} \cite{Murase2015} \cite{MuraseDermer2014}, galaxy clusters \cite{Fang}. 
 Among AGN, blazars are often considered the most probable candidate because of their jet pointed toward the line of sight to the Earth \cite{MuraseDermer2012}. These objects present peculiar characteristics such as variability at all frequencies and an intense emission in the $\gamma$-ray band \cite{Urry} \cite{Bottcher} \cite{Tavecchio2016}. This makes blazars the most numerous extragalactic $\gamma$-ray sources. Because of the beaming, the emission observed from blazars is dominated by the non-thermal continuum produced in the jet. This characterises the so-called spectral energy distribution, SED, typically showing a ``double hump'' shape. The low energy component, peaking between IR and soft-X rays,  is explained by the synchrotron radiation of relativistic electrons inside the jets, while the second component, usually peaking in $\gamma$-ray band, has not a completely clear origin. The most popular scenario is the leptonic model, where the second component is due to the Inverse Compton (IC) radiation from the same electrons producing the first component. In hadronic scenarios, instead, the second peak of SED is thought to originate from high-energy protons loosing energy through synchrotron emission or photo-meson reactions. Assuming a coexistence of both electrons and protons inside the jet, the favourite mechanism to produce high-energy neutrino from blazars is $p\gamma$ reaction.
 
Blazars are divided in two subclasses, flat spectrum radio quasar (FSRQ), characterised by broad emission lines typical of quasars, and BL Lacertae (BL Lac), showing extremely weak or absent emission lines in their optical spectra. FSRQs are generally more powerful than BL Lacs. The distinct difference between the two subclasses can be interpreted by a different nature of the the accretion flow \cite{Ghisellini2009} \cite{Ghisellini2017}.
At a first sight, FSRQ, characterised by intense thermal radiation, providing an ideal photon target field, seems the best blazar subclass to host $p\gamma$ reactions and produce neutrinos. Kadler et al.\cite{Kadler} found a coincidence in time between a long-lasting ($\sim$months) outburst of the FSRQ PKS B1414-418 and the arrival time of an HESE neutrino with an uncertainty region of $\sim$$16^\circ$.  However, there are several arguments against the possibility of FSRQs being the sources of the neutrinos revealed by IceCube. If FSRQs can produce neutrinos, the photon involved in the photo-meson reaction are most likely the UV photons of the broad line region (BLR). This implies a very-high energy $E_p$ of parent protons and thus of the neutrinos\footnote{For photo-meson reactions the approximate relation is $E_\nu \sim E_p/20$.}. Precisely, the energy of the proton must follow the photopion threshold condition $E_p\epsilon > m_\pi m_p c^4$ with $\epsilon$, the energy of the interacting photons, $m_\pi$ and $m_p$ the masses of pion and proton. Indeed, the spectra of neutrinos produced by FSRQs is predicted to be hard in the range of energy observed by IceCube \cite{Murase2015}, that instead reveals a relatively flat-soft spectrum. 

BL Lac objects seem disfavoured as $\nu$ emitters, mainly because their low luminosity hints to inefficient photo-meson production \cite{Murase2015}. However Tavecchio et al.\cite{Tavecchio2014} showed that if the jet is structured with a fast core (spine) and a slower layer, the neutrino emission from these object could match the observed intensity with an acceptable value of the cosmic ray power for the jet. This thesis is supported by Padovani et al.\cite{Padovani2016} which present the evidence for a significant spatial correlation between the reconstructed arrival direction of neutrinos (including both hemispheres, thus both HESE and through-going muon) and BL Lac objects emitting very high-energy $\gamma$-rays ($> 50$ GeV). 

\section{BL Lac objects as neutrinos emitters}
In Righi et al.\cite{Righi}, based on the results of [\cite{Padovani2016}] and [\cite{Tavecchio2014}], we selected a sample of high-energy emitting BL Lac (HBL) objects from the 2FHL catalogue and linked the emission of muon neutrinos coming from the photo-meson reaction to the $\gamma$-ray from Inverse Compton. We accept a leptonic scenario for the HBL electromagnetic emission; in this way we assume that any electromagnetic component associated to hadronic reactions, such as the decay of $\pi^0$ (and hence to neutrino emission) does not dominate the SED. We refer to Righi\cite{Righi} for a complete description of the scenario. Here we just recall the key points to find the linear relation between the bolometric neutrino flux $F_\nu$ for a given HBL source and its high-energy $\gamma$-ray flux $F_\gamma$.
The total, energy integrated, neutrino luminosity $L_\nu$ can be expressed as
$L_\nu=\epsilon_p Q_p \delta^4_s$;
where $\epsilon_p$ is the averaged efficiency for $\nu$ production, $Q_p$ the total cosmic ray injected power in the spine region and $\delta^4_s$ is the beaming factor determining the amplification of the emission always in the spine region. The high-energy $\gamma$-ray luminosity due to IC can be expressed in the same way but considering the relativistic electrons 
$L_\gamma=\epsilon_e Q_e \delta^4_s$; where $\epsilon_e$ measures the efficiency for $\gamma$-ray production.
Hence we have a relation of the ratio of the luminosities (and then of the fluxes):
\be
\frac{F_\nu}{F_\gamma}=\frac{L_\nu}{L_\gamma}=\frac{\epsilon_p Q_p}{\epsilon_e Q_e}
\ee
We assume that both efficiencies $\epsilon_p$ and $\epsilon_e$ depend on the same photon field (the layer radiation), and thus their ratio, $\epsilon_p/\epsilon_e$, is constant in first approximation. The same approximation could be done for $Q_p/Q_e$. In this case both can be linked to the total power carried by the jet $Q_{p,e}=\eta_{p,e}P_{jet}$ with the ratio $\eta_p/\eta_e \approx const$, hence
$F_\nu=k_{\nu\gamma}F_\gamma$.
In Righi et al.\cite{Righi} we derived the average value of $k_{\nu\gamma}$ comparing the total neutrino diffuse flux measured by IceCube and the entire high-energy $\gamma$-ray emission of HBL detected by Fermi.  With this calculation we were likely overestimating the neutrino flux for each sources because we didn't consider the unresolved HBL sources by Fermi. In fact we calculated the total $\gamma$-ray flux $F_\gamma$ by summing the fluxes of HBLs catalogued in the 2FHL\cite{Ackermann2016a} that includes all the sources detected above 50 GeV. It should be taken into account that, from the results by Ackermann et al.\cite{Ackermann2016b}, the derived neutrino fluxes could be lower by a factor $\approx 3$.
It further should be noted that in Righi et al.\cite{Righi} we used $k_{\nu\gamma}$ to calculate the neutrino flux for each HBL source, $F_{\nu_i}=k_{\nu\gamma}F_{\gamma_i}$, assuming that $k_{\nu\gamma}$ is the exactly same for all sources.

We can calculate the expected neutrino rate, $R_\nu$, in IceCube and Km3NeT for the brightest 2FHL HBL using the neutrino flux $F_{\nu_i}$ and the effective area of the instrument $A_{eff}$ (IceCube\cite{Yacobi} and Km3NeT\cite{Adrian-Martinez}):
\be
R_\nu= \int^{E_2}_{E_1} F_{\nu_i}(E_\nu)A_{eff}(E_\nu)dE_\nu
\ee 
where $T_{exp}$ is the integration time, one year in this case. The effective area $A_{eff}$ for IceCube is given in range of declinations ($0^\circ<\delta<30^\circ$,$30^\circ<\delta<60^\circ$,$60^\circ<\delta<90^\circ$) while for Km3NeT $A_{eff}$ is full-sky averaged. In table \ref{TableIce} we reported the main results of Righi et al. for IceCube and Km3NeT. We remark that we calculated the expected muon neutrino flux and the muon neutrino rate for each source because the good angular resolution ($\le 1^\circ$) of the through-going muon permits a possibile spatial correlation between the position of a source and the direction of the incoming neutrinos. 
For this reason, in the case of IceCube, we consider only the muon neutrinos coming from the northern hemisphere (this is the reason why for the last three sources in table \ref{TableIce} we do not report the expected rate number $R_\nu$ for IceCube). 
\begin{table}
\caption{Expected 0.1-10 PeV flux (in units of $10^{-8}$ GeV cm$^{-2}$ s$^{-1}$) and detection rate (yr$^{-1}$) of muon neutrino $\dot{N_{\nu}}$ for the brightest 2FHL BL Lacs with IceCube at different declinations (top) and with Km3NeT with the horizon as thresholds on the zenith angle (bottom).}
\footnotesize
\centering
\begin{tabular}{|l|c|c||c||c|c|}
\hline
& & \mco{1}{|c||}{IceCube} & \mco{1}{|c||}{Km3NeT}     & \mco{2}{|c|}{New approach}      \\
\hline
Name & $F_\nu$ & $R_{\nu}$  & $R_{\nu}$  &{$N_{\mu_{source}}$} & {$N_{\mu_{back}}$} \\
\hline
& &  & & &\\
 Mkn421                          &  8.77   &   4.89     &  4.59         & 10.23 & 8.46     \\ 
 PG1553+113                  &   1.89  &   2.47    &  1.42          & 4.00 & 8.87  \\    
 Mkn501                          &  3.41   &   1.90     &  1.65          & 3.86 & 8.44\\ 
  PKS1424+240               &   1.00  &  1.30     &  0.67          & 1.61 & 8.71\\  
   PG1218+304                &    0.92 &  1.20      &  0.55         & 1.27 & 8.60\\   
  TXS0518+211                &    0.87 &  1.14     &  0.59          & 1.48 & 8.74 \\   
 3C66A                             &  0.87   &  0.49     &  0.38          & 1.00 & 8.40\\
 PKS2155-304                  &  2.15  &       &  2.23                  & 3.00 & 8.60\\
& & & & &\\
\hline
\end{tabular}
\label{TableIce}
\end{table}
Our calculations predict that only for a few $\gamma$-ray bright HBL we expect a rate numbers $R_\nu$  detectable in few years of operation. For IceCube, in particular, there are only two sources, Mkn 421 and PG 1553+113, that show a rate exceeding 1 event yr$^{-1}$. For Mkn 421 we obtained a relatively large expected rate, 4.89 yr$^{-1}$. However, the declination of Mkn 421 is $+38^\circ$ $12'$ $31.7''$, close to the lower limit of declination range validity of the effective area ($30^\circ<\delta<60^\circ$). A finer binning of the effective area could be used to find a more precise expected neutrino rate number for the sources by IceCube. For Km3NeT instead we obtain an appreciable neutrino flux for several sources. While for IceCube the sources have the same visibility during the year because IceCube is located at South Pole, the analysis for Km3NeT is more complicated because the sources are partially visible (i.e. stays below the horizon) during the year. Km3NeT collaboration give the visibility as a function of source declination for the muon-track analysis for tracks below the horizon and up to $10^\circ$ above the horizon. Table \ref{TableIce} shows only the expected rate number $R_\nu$ for tracks below the horizon.

This work is missing of an analysis of the background due to the atmospheric neutrinos and an estimate of the sensitivity of a possible detection of the sources. 
Furthermore recently some arguments against BL Lac objects as candidates neutrino emitters have been raised. In particular Murase \& Waxman \cite{MuraseWaxman2016} presented an analysis of the constraints that can be put on the average luminosity and the local volume density of high-energy neutrino sources, based on the non-detection of multiplets in the detector (or, equivalently, on the non-detection of ``point sources" associated to high-energy neutrino-induced muon tracks). In this way they are able to rule out some of the possible source classes, in particular those characterised by a  large luminosity and a low cosmic density. Their calculation lead to exclude blazar (both FSRQ and BL Lac) as principal neutrino emitting source class. We note however that the BL Lac sample they consider as representative the BL Lacs population belonging to the 1FGL catalogue\cite{Ajello2014}. This sample includes all the BL Lac objects detected by \textit{Fermi}/LAT in the band $0.1$-$100$ GeV. In this way there is a selection effects disfavouring the HBL objects that have a lower flux in this band. Otherwise this subclass of BL Lac may be observed at higher energy range because of their second peak of SED is centred at high frequencies. Since HBL are more numerous and less powerful than the rest of BL Lac population, they could satisfy the constraints given by Murase \& Waxman. Also Vissani\cite{pall2} showed an analysis about HBL objects as candidates neutrino emitter.

For these reasons we thought to retake a calculation to obtain the expected muon rate (yr$^{-1}$) passing through the detector (we will consider only IceCube) without using the effective area of the detector, but performing a calculation starting on first principle. Although simplified, this approach provides an acceptable estimate\cite{Gaisser1990} \cite{Ribordy2012} \cite{MuraseWaxman2016} and, importantly, it allows one to calculate a significance of the possible detection considering also the background rate for each direction (and so for each source).

\section{A new approach}
The number of interactions ($\nu_\mu $X$ \rightarrow \mu$Y) per unit time $\dot{N}$ is given by the cross section $\sigma$ times the incident flux. In our case we consider only through-going events, and so $\nu_\mu$, because of their associated small angular uncertainty ($\le 1^\circ$).  In this context the number of muons per unit time crossing the detector is given by: 
 \begin{equation}
 d\dot{N}=F_\nu (E_\nu)e^{-\tau(x,E_\nu)}A \frac{\rho(x)}{m_p}\sigma_{CC}(E_\nu)dx
 \label{eq:N}
 \end{equation}
 where  $\sigma_{CC}$ is the cross section of charged current\footnote{We'll consider only the charged current interaction between $\nu_\mu N$.}, $F_\nu$ is the neutrino flux in GeV$^{-1}$ cm$^{-2}$ s$^{-1}$ sr$^{-1}$, $\tau$ takes into consideration the neutrino flux attenuation and depends on the path $x$, of the neutrino, inside the Earth (and so it depends on the zenith angle $\Theta$) and the energy of flux $E_\nu$; it corresponds to:
 \begin{equation}
 \tau(x,E_\nu)=\int_0^x \frac{\rho(x')}{m_p}\sigma_{CC}(E_\nu)d x'
 \end{equation}

The number of target nucleons per $dx$ is given by $A \frac{\rho(x)}{m_p}dx$, where $m_p$ is the mass of proton, $A$ is the detector projected area (which in principle depends on the zenith angle) that we approximate to $\simeq 1$km$^2$. $\rho$ is the Earth density (in g cm$^{-3}$) and depends on the path inside the Earth $x$.
Defining $dX=\rho(x)dx$ and dividing both members of equation \ref{eq:N} for $dX$ we obtain $d\dot{N}/dX$.
We want to study the rate of muon neutrino per energy $d\dot{N}/dE_\mu$, that it's equal to:
\begin{equation}
\frac{d\dot{N}}{dE_\mu}=\frac{d\dot{N}}{dX}\frac{dX}{dE_\mu}
\label{eq:dNdXdE}
\end{equation}
The first term right of equation \ref{eq:dNdXdE} is given by equation \ref{eq:N}, while the second term derives from the inverse of the average muon energy-loss rate
$
-\frac{dE_\mu}{dX}=\alpha+\beta E_\mu
$
where $\alpha$ is the ionization term while $\beta E_\mu$ is the radiative term at TeV range $\alpha$ and $\beta$ are respectively equal to $\simeq 2 \cdot 10^{-3}$ GeV cm$^2$ g$^{-1}$ and  $\simeq 5 \cdot 10^{-6}$ cm$^2$ g$^{-1}$.
Replacing previous equations in eq.\ref{eq:dNdXdE} we obtain:
\begin{equation}
\frac{d\dot{N}}{dE_\mu}=\frac{1}{\alpha+\beta E_\mu}\frac{A}{m_p}\int_{E_\mu}^{E_{\nu_{max}}} F_\nu(E_\nu) \sigma_{CC}(E_\nu)e^{-\tau(x,E_\nu)}dE_\nu
\label{eq:finale}
\end{equation}
 We have to integrate between the minimum and maximum value of the incoming neutrino to produce a muon of energy $E_\mu$. The minimum value of neutrino energy to produce a muon with energy $E_\mu$ corresponds to $E_{\nu_{min}}=E_\mu$  corresponding to neutrinos with energy $E_\mu$ interacting just before the detector.
 
Integrating for all possible $E_\mu$, we obtain the number of muon (or neutrino) produced in a time $T$ for every source:
 \begin{equation}
 N=\varepsilon T \int_{E_{\mu_{min}}}^{E_{\mu_{max}}} \frac{d\dot N}{dE_\mu}dE_\mu
 \label{eq:Nfin}
 \end{equation}
 where $\varepsilon \leqslant 1$ is the efficiency of the detector. 
 We use equation \ref{eq:Nfin} to calculate the expected muon rate coming from the HBL objects of 2FHL catalog and the background rate for each source. We use the cross section $\sigma_{CC}(E_\nu)$ of eq.\ref{eq:finale} given in Connolly et al. \cite{Connolly2011}, the neutrino flux $F_\nu(E_\nu)$ found in Righi et al. and the Earth density $\rho(x)$ reported by Dziewonski \& Anderson\cite{Dziewonski}. Main backgrounds to the search for astrophysical neutrinos are high-energy atmospheric neutrinos and muons produced by cosmic-ray interactions in the Earth's atmosphere. There are two atmospherical neutrinos components: the conventional neutrinos and the prompt neutrinos produced in the atmosphere by the decay of charmed particles.
To find the number of background muon produced in a time $T$ at the same declination angle of the BL Lac sources, we have to consider the dependence on the solid angle $d\Omega$. For this reason the background flux to use in eq.\ref{eq:finale} is 
\be
F(E_\nu)_{Back}=2\pi \int_0^\Theta \phi_B(\Omega, E_\nu) d\Omega \approx \pi \Theta^2 \phi_B(E_\nu) 
\ee
where we consider $\Theta=1^\circ$ and $\phi_B(E_\nu)$ the flux of background given by Aartsen (2016) in GeV$^{-1}$ cm$^{-2}$ s$^{-1}$ sr$^{-1}$. The energy range of background neutrinos is $10^2$ GeV $<E_{\nu_{back}}<10^6$ GeV.

 In this procedure we do not include a detailed analysis of the efficiency of the detector. For this reason the expected muon rates N will be likely overestimated.
 
 Last two columns of table \ref{TableIce} show the expected muon rate produced by muon neutrinos of the sources and the background passing through the detector.  
These numbers are subject to stochastic fluctuations, for this reason they need to be treated with Poisson distribution. 
Li and Ma\cite{Li1983} gives a formulae to estimate the significance $S$ of observations.

\begin{figure}
\centerline{\includegraphics[width=0.50\linewidth]{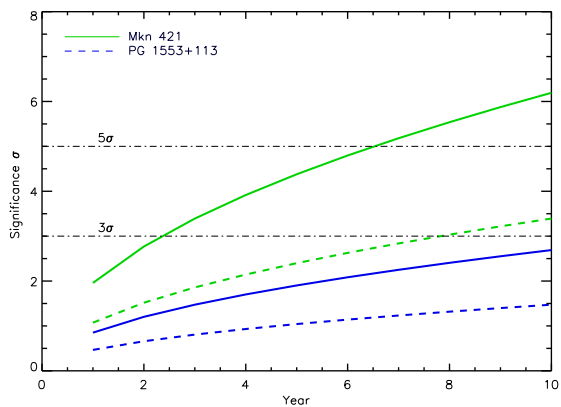}}
\caption[]{same figure with draft option (left), normal (center) and rotated (right)}
\label{fig:fin}
\end{figure}

We assume a Poisson distribution around the number of muon for the sources Mkn 421 and PG 1553+113 and the corresponding background; we extract randomly a value from the Poisson distributions and we calculate the significance $S$. We repeat this procedure 10000 times, in this way we obtain a distribution around the most probable significance for the two sources.
Figures \ref{fig:fin} shows the position of the peak of the distribution for Mkn 421 and PG 1553+113. Solid line consider the efficiency $\varepsilon=1$, while dashed lines consider an efficiency of the instrument of 30\%, $\varepsilon=0.3$.

\end{document}